\documentclass[fleqn,usenatbib]{mnras}

\usepackage{newtxtext,newtxmath}

\usepackage[T1]{fontenc}

\DeclareRobustCommand{\VAN}[3]{#2}
\let\VANthebibliography\thebibliography
\def\thebibliography{\DeclareRobustCommand{\VAN}[3]{##3}\VANthebibliography}

\usepackage{graphicx}	
\usepackage{amsmath}	
\usepackage{float}

\title[Fragmentation induced starvation in Population III star formation: a resolution study]{Fragmentation induced starvation in Population III star formation: a resolution study}

\author[L. R. Prole]{
Lewis R. Prole,$^{1}$\thanks{E-mail: Prolel@cardiff.ac.uk}
Paul C. Clark,$^{1}$
Ralf S. Klessen,$^{2,3}$
Simon C. O. Glover$^{2}$
\\
$^{1}$Cardiff University School of Physics and Astronomy\\
$^{2}$Zentrum f{\"u}r Astronomie, Universit{\"a}t Heidelberg, Institut f{\"u}r Theoretische Astrophysik, Albert-Ueberle-Str. 2, 69120 Heidelberg, Germany\\
$^{3}$Interdisziplin{\"a}res Zentrum fur Wissenschaftliches Rechnen, INF 205, D-69120, Heidelberg, Germany
}

\date{Accepted XXX. Received YYY; in original form ZZZ}

\pubyear{2020}

\begin{document}

\label{firstpage}
\pagerange{\pageref{firstpage}--\pageref{lastpage}}
\maketitle

\begin{abstract}
The Population III initial mass function (IMF) is currently unknown, but recent studies agree that fragmentation of primordial gas gives a broader IMF than the initially suggested singular star per halo. In this study we introduce sink particle mergers into {\sc Arepo}, to perform the first resolution study for primordial star formation simulations and present the first Population III simulations to run up to densities of 10$^{-6}$g cm$^{-3}$ for hundreds of years after the formation of sink particles. The total number of sinks formed increases with increasing sink particle creation density, without achieving numerical convergence. The total mass in sinks remains invariant to the maximum resolution and is safe to estimate using low resolution studies. This results in an IMF that shifts towards lower masses with increasing resolution. Greater numbers of sinks cause increased fragmentation-induced starvation of the most massive sink, yielding lower accretion rates, masses and ionising photons emitted per second. The lack of convergence up to densities 2 orders of magnitudes higher than all relevant chemical reactions suggests that the number of sinks will continue to grow with increasing resolution until H$_{2}$ is fully dissociated and the collapse becomes almost adiabatic at 10$^{-4}$g cm$^{-3}$. These results imply that many Population III studies utilising sink particles have produced IMFs which have overestimated the masses of primordial stars, and underestimated the number of stars formed. In the highest resolution runs, sinks with masses capable of surviving until the present day had an ejection fraction of 0.21.

\end{abstract}

\begin{keywords}
stars: Population III -- stars: formation -- hydrodynamics -- stars: luminosity function, mass function -- software: simulations
\end{keywords}



\section{Introduction}
The first stars, known as Population III (Pop III) stars, are responsible for the first ionising radiation, which began the epoch of re-ionisation \citep{Bromm2001}. When they died as supernovae, they injected the interstellar medium (ISM) with the first metals \citep{Heger2003}, which were then incorporated into the next generation of stars (Population II).

 As the amount of ionising radiation a star emits and its eventual fate is dependent on its mass, the initial mass function (IMF) of Pop III stars has a huge effect on the evolution of the Universe. Initially it was thought that Pop III stars formed in isolation \citep{Haiman1996}, and were massive \citep{Abel2002,Bromm2002}, yet further studies showed that clouds forming Pop III stars were susceptible to fragmentation in the presence of subsonic turbulence \citep{Clark2011}. Since then, numerical studies have attempted to improve the picture of Pop III star formation by including feedback mechanisms \citep{OShea2008}, live dark matter potentials \citep{Stacy2014} and magnetic fields (e.g. \citealt{Machida2008a, Machida2013, Sharda2020}). Despite this, the Pop III IMF is still in dispute, and there are still many factors left to study. 
 
Existing attempts to produce the primordial IMF differ considerably in terms of the maximum density their simulations run to, which is directly linked to the maximum resolution of the simulation. The Jeans length $\lambda_{\text{J}}$ of a structure of given density and temperature marks the maximum size it can achieve before thermal pressure cannot prevent gravitational collapse. Artificial fragmentation occurs in hydrodynamic codes if the local $\lambda_{\text{J}}$ falls below the size of mesh cells $\Delta x$. The Truelove condition \citep{Truelove1997} requires a Jeans number $\Delta x/\lambda_{\text{J}}$ of 0.25 or less, corresponding to at least 4 cells spanning across any $\lambda_{\text{J}}$, to prevent artificial fragmentation. This requirement is increased to 30 cells per Jeans length for magneto-hydrodynamical (MHD) simulations, in order to capture dynamo amplification of the magnetic field \citep{Federrath2011}. Adaptive mesh refinement (AMR: \citealt{Berger1989}) codes can refine the mesh based on the local $\lambda_{\text{J}}$ to meet this resolution criteria.  

Another condition for numerical stability is known as the Courant condition \citep{Courant1952}, which states that information must not be allowed to travel further than the cell length during a timestep, such that information from a cell can only be communicated to its immediate neighbours. As a result, the timestep has to decrease as the mesh becomes more refined.

Numerical simulations cannot refine indefinitely. As the gas gets denser (decreasing $\lambda_{\text{J}}$), the number of cells increases. These new cells are smaller and require smaller timesteps to satisfy the Courant condition, making it increasingly computationally expensive to run to higher densities. Sink particles \citep{Bate1995,Federrath2010} provide an alternative to indefinite refinement. They are non-gaseous particles that contain all of the mass within the volume they occupy and can accrete matter from their surrounding cells. As they cannot fragment -- either naturally or artificially -- their implementation at high densities overcomes the problem posed by the Jeans refinement criterion.

In present day star formation simulations, the sink creation density is chosen to be that of the first adiabatic core. As laid out in \cite{Larson1969}, the initial isothermal collapse of a cloud is halted in the central region when the gas becomes opaque to outgoing radiation. At $\rho \sim 10^{-10}$~g~cm$^{-3}$, the central temperature and density are such that collapse stops in the central region, forming the first adiabatic core, while the material outside the core continues to freefall isothermally. At this point the core is stable to further fragmentation. The radial density profile inside the core is flat and extends out to $\lambda_{\text{J}}$, so the radius of the of sink particle is chosen to be the Jeans length at the creation density and temperature. 

The most important source of the opacity that halts collapse and prevents fragmentation in the first core is dust grains \citep{Low1976}, which are not present in primordial star formation. In this regime, there is no clear `first core', as is apparent from the thermal evolution described in e.g.~\cite{Omukai2005}. Referring to fig.\ref{fig:simple}, primordial gas is cooled to $\sim$200 K at $\sim$10$^{-20}$g cm$^{-3}$ by molecular hydrogen cooling (A). Gravitational collapse allows the gas temperature to rise from 200 K to 1000 K by 10$^{-15}$g cm$^{-3}$ (B). Here three-body reactions convert most of the hydrogen into molecules. At 10$^{-12}$g cm$^{-3}$, the H$_2$ cooling rate decreases due to  gas opacity (C), but collision-induced emission kicks in to become the dominant cooling process at 10$^{-10}$g cm$^{-3}$ (D). Once the temperature reaches 2000 K at 10$^{-8}$g cm$^{-3}$, dissociation of H$_2$ molecules provides effective cooling (E) until it is depleted and the collapse becomes almost adiabatic at 10$^{-4}$g cm$^{-3}$ (F). 

Running simulations up to this density with a full chemical treatment lies beyond current computational capabilities. The appropriate density to replace gas with a stable sink particle is unclear, but the gas must be stable to fragmentation.  \cite{Gammie2001} suggested that discs become stable to fragmentation when heating from turbulence is balanced by cooling. \cite{Smith2011} studied the stability of Pop III collapses against fragmentation during the early period of protostar formation, before ionising radiation from the star becomes important, where accretion luminosity feedback is the main opposition to fragmentation. They found that accretion luminosity delays fragmentation but does not prevent it. 

The sink particle creation density varies between different studies. For example, \citet{Stacy2010}, \cite{Stacy2012}, \cite{Clark2011}, \citet{Susa2014} and \cite{Sharda2020} all introduce sinks at densities $\sim 10^{-12}$--$10^{-11} \: {\rm g \, cm^{-3}}$, in analogy with present day star formation, while \cite{Smith2011}, \citet{Stacy2016} and \cite{Wollenberg2019} introduce them at $\sim$10$^{-9}$g cm$^{-3}$, above which there are no chemical heating terms that can prevent the gas from collapsing. Among the highest resolution Pop III studies, \cite{Greif2011}, \cite{Clark2011a} and \citet{Hartwig2015a} introduce their sink particles at $\sim$10$^{-7}$g cm$^{-3}$. 

Sink particles are not a perfect solution to the indefinite refinement problem, and an author's choice of sink particle creation density may change the morphology of the resulting cluster.
Therefore, some authors have used other approaches to address this problem. For example, \cite{Greif2012} ran several simulations without sinks by simply following the collapse of the gas up to the density of $10^{-4} \: {\rm g \, cm^{-3}}$ at which the further collapse becomes adiabatic.
The disadvantage of this approach is that despite a considerable investment of computing time, the simulations could only follow the first 10~yr of the evolution of the system after the formation of the first protostar. \cite{Hirano2017} avoided this difficulty by artificially suppressing cooling above a pre-chosen threshold density, resulting in the gas evolving adiabatically above this density. They examined several different threshold densities ranging from $\sim 10^{-14} \: {\rm g \, cm^{-3}}$ to $\sim 10^{-9} \: {\rm g \, cm^{-3}}$, and showed that this allowed them to run the simulations for periods of several hundred years (in the high threshold case) to several tens of thousands of years (in the low threshold case) after the formation of the first protostar. However, a disadvantage of this suppressed-cooling approach is that the pressure-supported clumps that form in these simulations have sizes that are much larger than those of real Pop III protostars. For example, the clumps formed in the simulation with the highest threshold density have sizes of $\sim 10$--60~AU, much larger than the physical size of $\sim 0.5$~AU that we expect for massive pre-main sequence Pop III stars. How well the behaviour of these artificially large clumps represents the behaviour of real Pop III protostars remains unclear.
In view of these limitations of current sink-free approaches, sink particles remain an essential tool for exploring Population III star formation past the point of the initial collapse and assessing the resulting primordial IMF.

In this paper, we perform the fist Pop III resolution test, exploring the effects of increasing the maximum resolution of primordial star formation simulations by varying the sink particle creation density $\rho_{\text{sink}}$ from 10$^{-10}$ -- 10$^{-6}$g cm$^{-3}$. We present the highest resolution Pop III simulations to date to have followed the system for hundreds of years after the formation of the first sink.  We examine the impact that this has on the fragmentation of Pop III accretion discs and the subsequent accretion rates onto the newly-formed protostars. Although primordial magnetic fields are believed to be present during Pop III star formation (e.g. \citealt{Federrath2011, Schober2015}), here we focus on a purely hydrodynamical scenario for simplicity, deferring an examination of the magnetohydrodynamical case to future work. We show that the number of sinks formed increases with $\rho_{\text{sink}}$ over the whole range tested, causing a considerable increase in fragmentation-induced starvation of the central star.

In section \ref{sec:method}, we describe the numerical set up of the simulations, chemistry and cooling model, use of sink particles, initial conditions and the newly implemented sink merger routine. In section \ref{sec:fragmentation} we analyse the fragmentation behaviour of the primordial gas along with the IMFs produced, showing that increased resolution results in increasing fragmentation-induced starvation. In section \ref{sec:ejections}, we describe our method of detecting ejections from the system and present the ejection fractions, noting that a considerable number of stars with masses allowing survival until the present day are ejected from the system. In section \ref{sec:discussion}, we summarise the results and speculate how the fragmentation-induced starvation changes the star's main sequence mass and the ionising effect on its surroundings. We also discuss the ability of current and future archaeological surveys to detect ejected Pop III stars. In section \ref{sec:caveats} we discuss caveats before concluding in section \ref{sec:conclusion}.

\section{Numerical method}
\label{sec:method}
\subsection{{\sc Arepo}}
We use the moving mesh code {\sc Arepo} \citep{Springel2010} to perform turbulent cloud collapses with a primordial chemistry set-up. {\sc Arepo} combines the advantages of AMR and smoothed particle hydrodynamics (SPH: \citealt{Monaghan1992}) with a mesh made up of a moving, unstructured, Voronoi tessellation of discrete points. {\sc Arepo} solves hyperbolic conservation laws of ideal hydrodynamics with a finite volume approach, based on a second-order unsplit Godunov scheme with an exact Riemann solver. Automatic and continuous refinement overcome the challenge of structure growth associated with AMR (e.g. \citealt{Heitmann2008}).

\subsection{Chemistry}
\label{chem}
We use the same chemistry and cooling as \cite{Wollenberg2019}, which is based on the fully time-dependent chemical network described in the appendix of \cite{Clark2011}, but with updated rate coefficients, as summarised in \cite{Schauer2017}. The network has 45 chemical reactions to model primordial gas made up of 12 species: H, H$^{+}$, H$^{-}$, H$^{+}_{2}$ , H$_{2}$, He, He$^{+}$, He$^{++}$, D, D$^{+}$, HD and free electrons.  Included in the network are: H$_{2}$ cooling (including an approximate treatment of the effects of opacity), collisionally-induced H$_{2}$ emission, HD cooling, ionisation and recombination, heating and cooling from changes in the chemical make-up of the gas and from shocks, compression and expansion of the gas, three-body H$_{2}$ formation and heating from accretion luminosity. For reasons of computational efficiency, the network switches off tracking of deuterium chemistry\footnote{Note that HD cooling continues to be included in the model.} at densities above 10$^{-16}$~g~cm$^{-3}$, instead assuming that the ratio of HD to H$_{2}$ at these densities is given by the cosmological D to H ratio of 2.6$\times$10$^{-5}$. The adiabatic index of the gas is computed as a function of chemical composition and temperature with the {\sc Arepo} HLLD Riemann solver.

\subsection{Sink particles}
 Sink particles are point masses inserted into the mesh to prevent artificial collapse when the local Jeans length falls below the minimum cell size of the mesh. The sink particle implementation we use in {\sc Arepo} was introduced in \cite{Wollenberg2019} and \citet{Tress2020}. Briefly, a cell is converted into a sink particle if it satisfies three criteria: 1) it reaches a threshold density; 2) it is sufficiently far away from pre-existing sink particles so that their accretion radii do not overlap; 3) the gas occupying the region inside the sink is gravitationally bound and collapsing. Likewise, for the sink particle to accrete mass from surrounding cells it must meet two criteria: 1) the cell lies within the accretion radius; 2) it is gravitationally bound to the sink particle. A sink particle can accrete up to 90$\%$ of a cell's mass, above which the cell is removed and the total cell mass is transferred to the sink.

The sink particle treatment also includes the accretion luminosity feedback from \citep{Smith2011}. However, the internal luminosity of the star is not included in this work, so our simulations are only valid until a time when the core is expected to begin propagating its accumulated heat as a luminosity wave.

\begin{figure}
	\hbox{\hspace{-0.5cm} \includegraphics[scale=0.6]{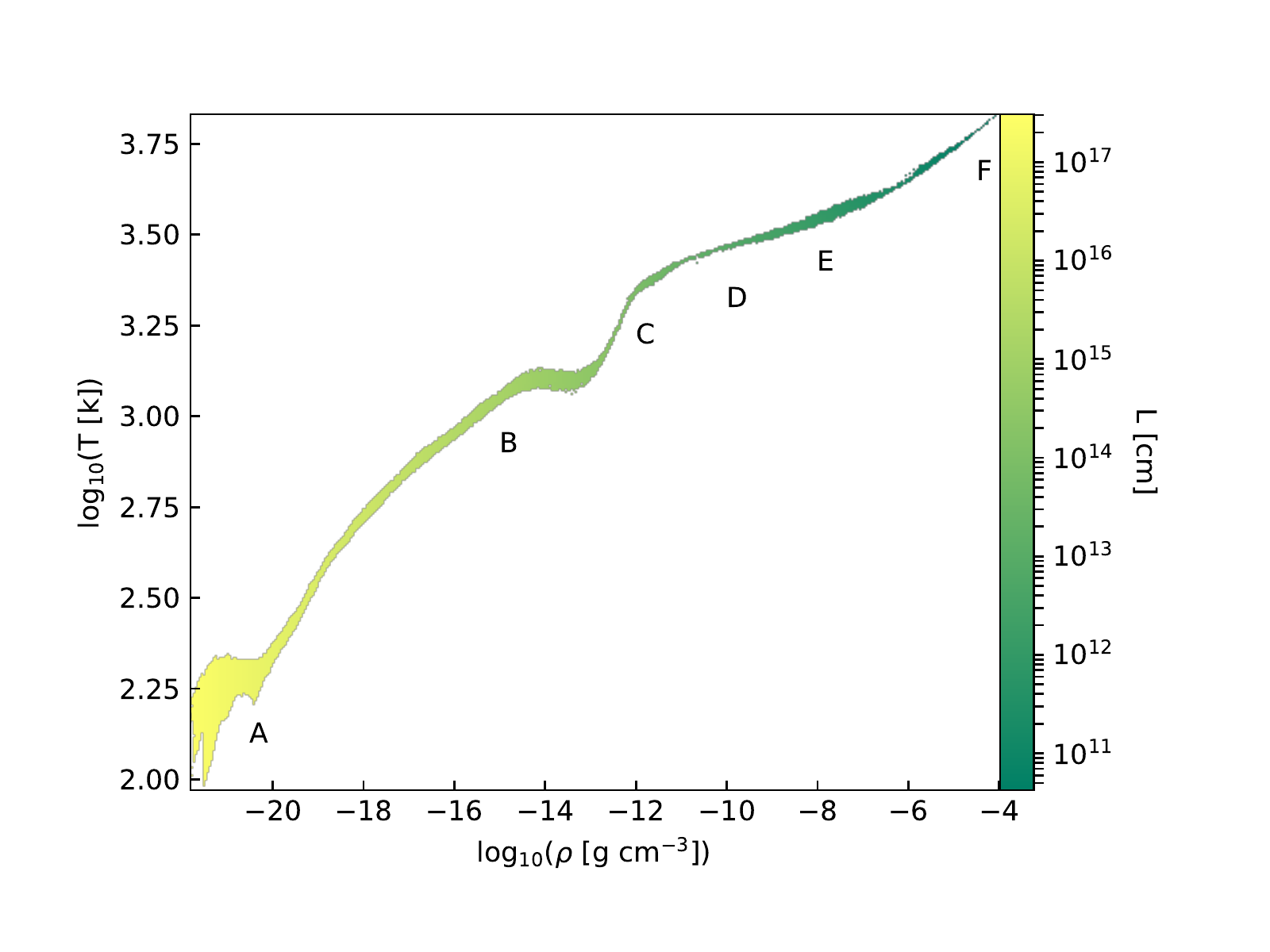}}
    \caption{Relationship between density and temperature during a purely gravitational collapse, resulting in a single central dense object. The temperatures were used to calculate the parameters in Table~\ref{table:1}. Shown in colour are the cell lengths as dictated by the chosen refinement criterion of 16 cells per Jeans length.}
    \label{fig:simple}
\end{figure}

The accretion radius of a sink particle $R_{\text{sink}}$ is chosen to be $\lambda_{\text{J}}$ corresponding to the sink creation density, given by
\begin{equation}
    \lambda_{\text{J}}=\sqrt{  \frac{k_B T} {G \rho_{\rm sink} (\mu m_{\rm p})}}.
	\label{eq:jeans}
\end{equation}
where $k_B$ is the Boltzmann constant, T is the temperature, $\rho_{\text{sink}}$ is the sink creation density, $\mu$ is the mean molecular weight and $m_{\rm p}$ is the mass of a proton. To estimate $\lambda_{\text{J}}$ before running the simulation, an realistic value of T at $\rho_{\text{sink}}$ is needed. To achieve this, a non-turbulent version of the simulations described in section \ref{Sims} was run, resulting in a single star. The simulation was run up until the maximum density reached 10$^{-4}$g cm$^{-3}$. fig. \ref{fig:simple} shows the resulting relationship between density and temperature. This gives an effective relationship between $\rho$ and $\lambda_{\text{J}}$ using Equation~\ref{eq:jeans}. The sink radius $R_{\text{sink}}$ is given by $\lambda_{\text{J}}$. We set the minimum cell length to fit 16 cells across the sink particle in compliance with the Truelove condition, giving a minimum cell volume $V_{\text{min}}=(R_{\text{sink}}/8)^3$. The minimum gravitational softening length for cells and sink particles $L_{\text{soft}}$ is set to $R_{\text{sink}}/8$. The values of $\rho_{\text{sink}}$, T, $R_{\text{sink}}$, minimum cell volume and minimum gravitational softening lengths are given in Table~\ref{table:1}.

\begin{table}
	\centering
	\caption{Sink creation density, temperature (fig. \ref{fig:simple}), sink radius, minimum cell volume and minimum gravitational softening lengths used in the study.}
	\label{table:1}
	\begin{tabular}{lcccr} 
		\hline
		$\rho_{\text{sink}}$ [g cm$^{-3}$] & T [K] & $R_{\text{sink}}$ [cm] & $V_{\text{min}}$ [cm$^{3}$] & $L_{\text{soft}}$ [cm]\\
		\hline
		$10^{-10}$ & 3050 & 1.37$\times 10^{14}$ & 5.10$\times 10^{39}$ & 1.72$\times 10^{13}$\\
		$10^{-9}$ & 3350 & 4.56$\times 10^{13}$ & 1.86$\times 10^{38}$ & 5.70$\times 10^{12}$\\
		$10^{-8}$ & 3750 & 1.53$\times 10^{13}$ & 6.95$\times 10^{36}$ & 1.91$\times 10^{12}$\\
		$10^{-7}$ & 4100 & 5.05$\times 10^{12}$ & 2.51$\times 10^{35}$ & 6.31$\times 10^{11}$\\
		$10^{-6}$ & 4460 & 1.67$\times 10^{12}$  & 9.03$\times 10^{33}$ &  2.08$\times 10^{11}$\\
		\hline
	\end{tabular}
\end{table}

\subsection{Initial conditions}
\label{Sims}
Primordial cloud collapse simulations were performed for 3 initial turbulent velocity seed fields (hereafter referred to as A, B and C) and 2 initial ratios of kinetic to gravitational energies $\alpha$. A resolution study was performed for the lower $\alpha$ set-up, where we ran simulations with 5 different values for the sink particle creation density (see Table~\ref{table:1}) for each of the different seed fields. In these simulations, the rms turbulent velocity was scaled to give a ratio of kinetic to gravitational energy of $\alpha=0.05$, to give a weak velocity field which encourages a more straightforward collapse onto a central object. We also performed higher velocity ($\alpha=0.25$) versions of the highest resolution runs, giving 18 simulations in total. We generate random velocity fields from the turbulent power spectrum $ P(k) \propto k^{-2}$. The initial conditions are similar to those of \cite{Wollenberg2019}, consisting of a stable Bonner Ebert sphere \citep{Ebert1955,Bonnor1956}, categorised by radius $R_{\text{BE}}=1.87$~pc and central density $\rho_c=2 \times$10$^{-20}$g cm$^{-3}$, which was enhanced by a factor of 1.87 to promote collapse, giving a total mass of 2720M$_\odot$ enclosed. The sphere was placed in a box of side length 4$R_{\text{BE}}$ with an initial temperature of 200 K and the simulations were performed with refinement criteria of 16 cells per Jeans length. The chemistry used was the same as \cite{Clark2011}, with initial abundances for H$_2$, H$^{+}$, D$^{+}$ and HD of $x_{\text{H}_{2}}=10^{-3}$, $x_{\text{H}^{+}}=10^{-7}$, $x_{\text{D}^{+}}=2.6\times$10$^{-12}$ and $x_{\text{HD}}=3\times$10$^{-7}$, respectively.

\subsection{Sink mergers}
Previous Pop III studies have shown that merging of dense objects is important to the outcome of the system. \cite{Greif2012} found that about half of the secondary protostars which formed in the disc migrated to the centre of the cloud in a free-fall time, where they merge with the primary protostar, boosting its mass to five times higher than the second most massive protostar. \cite{Hirano2017} found that fragments efficiently migrate to the primary protostar until the central star evacuates the surrounding gas by its radiation. Fragments large enough to form black holes may even merge generating an observable gravitational wave signal. More reecently \cite{Susa2019} noted a rapid merging phase once the disc fragments.

The total number of sinks and their masses would be unrepresentative of the IMF if they are not able to merge with one another. Similarly to \cite{Federrath2010}, we allow sinks to merge if they fulfil four criteria: 1) they lie within each other's accretion radius;  2) they are moving towards each other; 3) their relative accelerations are $<0$; and 4) they are gravitationally bound to each other. Since sink particles carry no thermal data, the last criteria simply requires that their gravitational potential well exceeds the kinetic energy of the system. When these criteria are met, the larger of the sinks gains the mass and linear momentum of smaller sink, and its position is shifted to the center of mass of the system. We allow multiple mergers per time-step, based on mass hierarchy; for example, if sink A is flagged to merge into sink B, and sink B is flagged to merge into sink C, then both A and B will be merged into sink C simultaneously.

\section{Fragmentation behaviour}
\label{sec:fragmentation}
\begin{figure*}
	\hbox{\hspace{-2cm} \includegraphics[scale=0.9]{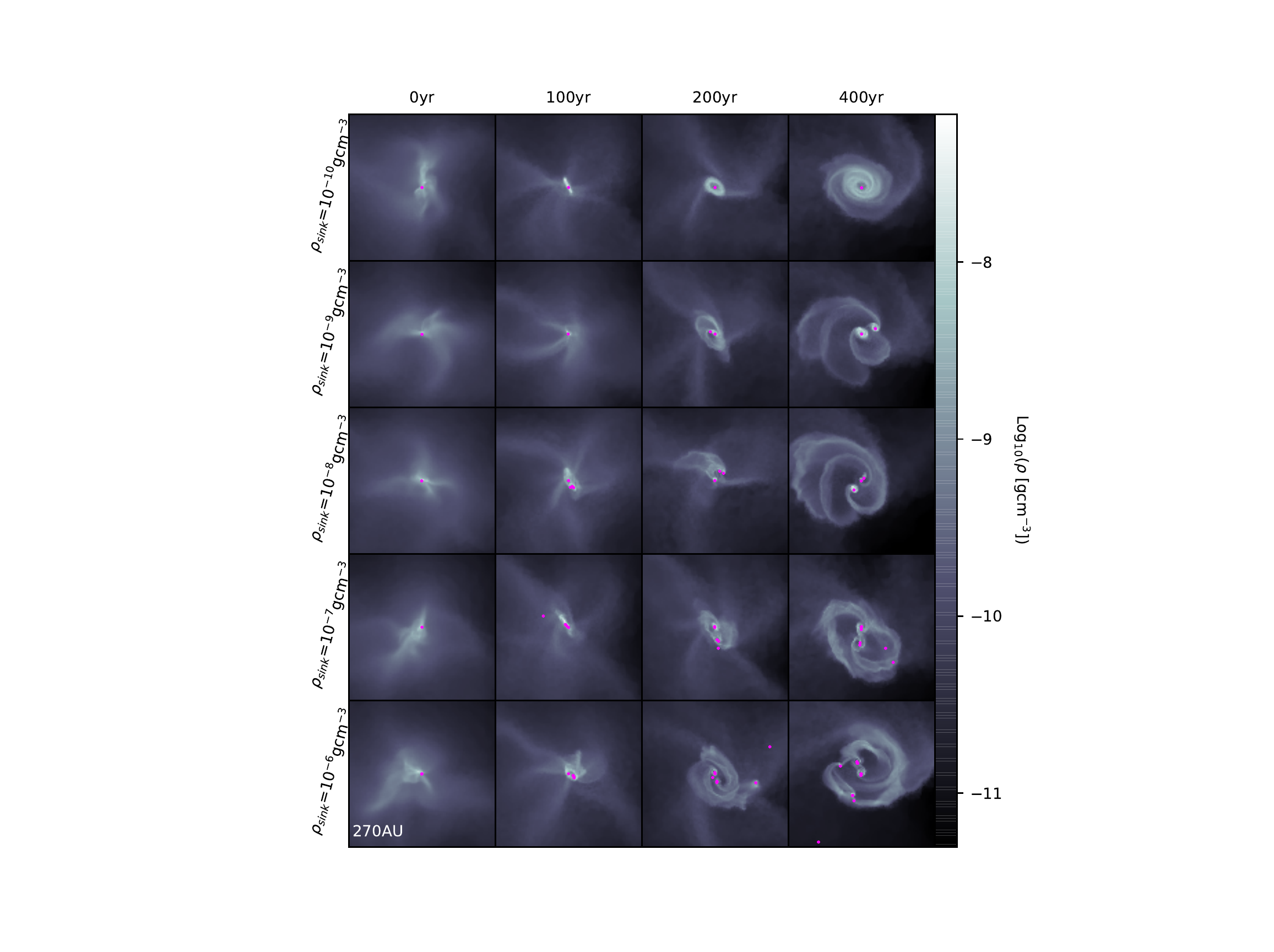}}
    \caption{For initial velocity field A, the inner 270~AU of the {\sc Arepo} unstructured density distributions projected onto uniform 500$^3$ grid cubes, flattened by summing the density over the $z$-axis. From left to right, we show
    snapshots taken at 0, 100, 200 and 400 years after the formation of the first sink particle, 
    for sink creation densities ranging from
    $rho_{\rm sink} = 10^{-10} \: {\rm g \: cm^{-3}}$ (top row) to $10^{-6} \: {\rm g \: cm^{-3}}$ (bottom row). Sink particles are shown as magenta dots. }
    \label{fig:grid}
\end{figure*}

Density projections for seed field A are shown at 0, 100, 200 and 400 years after the formation of the first sink in fig.\ref{fig:grid}. The increase in structure and fragmentation is clear as the simulation resolution is increased. Projections of seeds B and C at 400yr are given in Appendix A. fig.\ref{fig:sinks} shows the evolution of $N_{\text{sink}}$ and $M_{\text{tot}}$ with time in the simulations making up our resolution study. We see that for all three seed fields, the total number of sinks formed increases with increasing $\rho_{\rm sink}$ without achieving numerical convergence. 

Higher maximum gas density lowers the minimum Jeans scale and allows gravitational instabilities on smaller spacial scales, resulting in more fragmentation of the gas. As there are no new chemical reactions introduced for densities above 10$^{-8}$g cm$^{-3}$, this implies that fragmentation will continue to increase with increasing resolution until the gas becomes almost adiabatic at 10$^{-4}$g cm$^{-3}$, at which point it should be stable to fragmentation assuming the parallels with present day star formation hold. This means that all Pop III studies to date have underestimated the number of stars formed in their halos. However, the total mass of the system is unaffected by increased resolution, so estimates of the mass in stars within dark matter halos can be made with fairly low resolution simulations.

\begin{figure*}
	\hbox{\hspace{0.5cm} \includegraphics[scale=0.7]{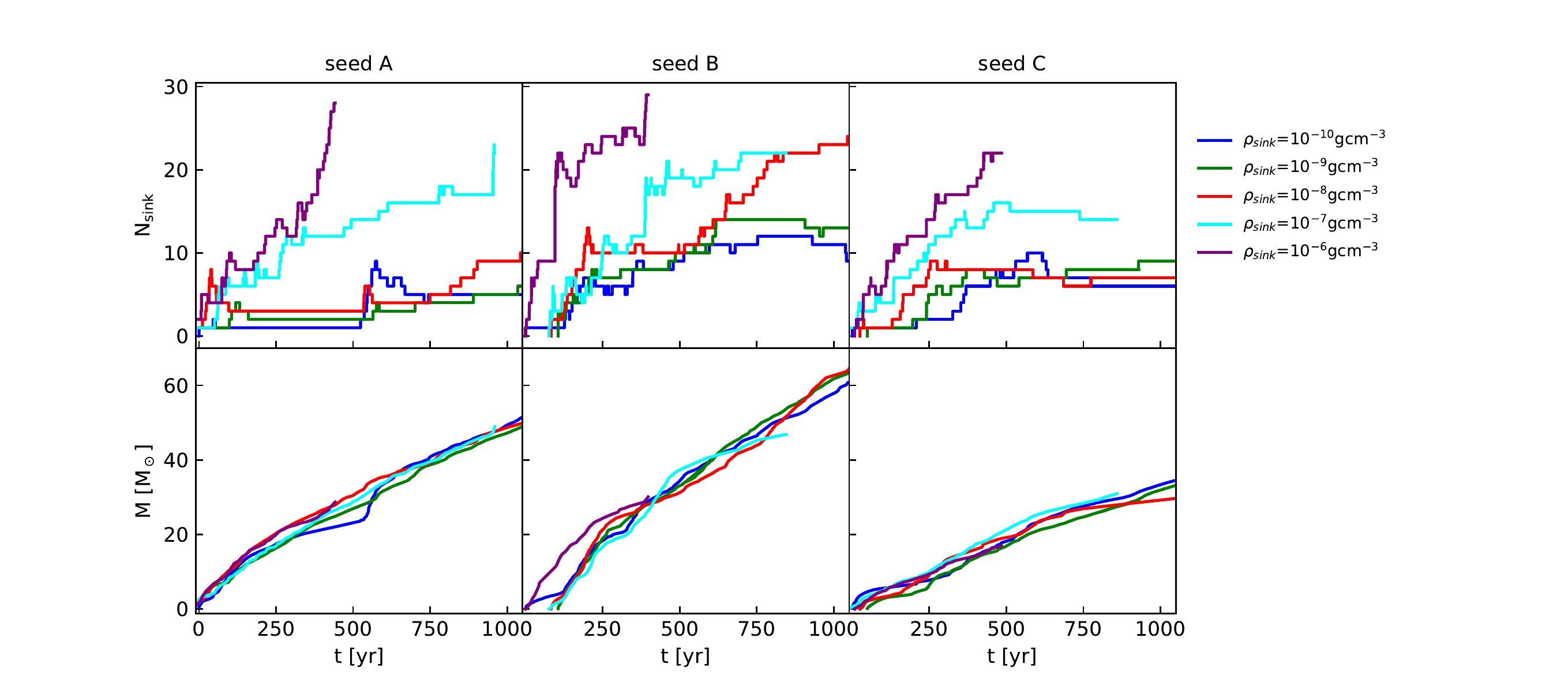}}
    \caption{Time evolution of the number of sinks ($N_{\rm sink}$) and the total mass $M$ in sinks for $\rho_{\text{sink}}=10^{-10}$-10$^{-6}$g cm$^{-3}$. Note that decreasing $N_{\rm sink}$ indicates sink merging.}
    \label{fig:sinks}
\end{figure*}
\begin{figure*}
	\hbox{\hspace{-1.5cm} \includegraphics[scale=0.7]{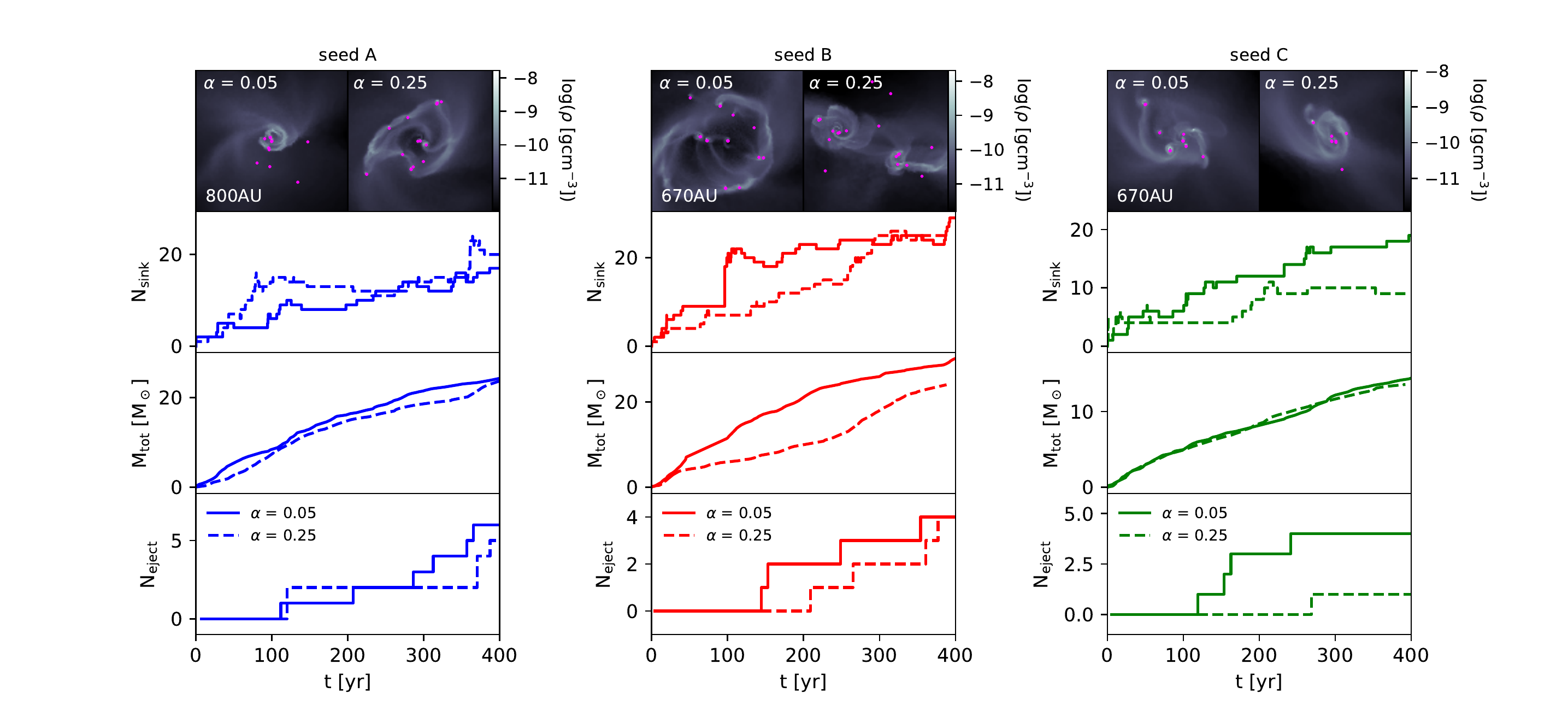}}
    \caption{Top to bottom: Comparisons of the system for $\alpha=0.05$ (left) and $\alpha=0.25$ (right) at 400~yr after the formation of the first sink. We compare the number of sinks formed, the total mass in sinks, and the number of sinks ejected from the system as a function of time, for $\alpha=0.05$ (solid lines) and $\alpha=0.25$ (dashed lines).}
    \label{fig:vels}
\end{figure*}


fig.\ref{fig:vels} compares the outcome of the $\alpha=0.05$, $\rho_{\rm sink} = 10^{-6}$~g~cm$^{-3}$ simulations with that of the corresponding runs with $\alpha=0.25$. Seed field B shows that the structure of the systems can be drastically different (likely due to the sudden burst in star formation at 100~yr in the low velocity case), while seed C shows that the ejection history can change. Despite these differences, it is clear that the high degree of fragmentation seen in the low velocity runs remains when the velocity fields are boosted to higher strengths. This demonstrates that the lack of convergence in the fragmentation in the resolution test was not a product of our choice of $\alpha$, but rather is a general result.

\begin{figure}
	 \hbox{\hspace{-0.5cm} \includegraphics[scale=0.6]{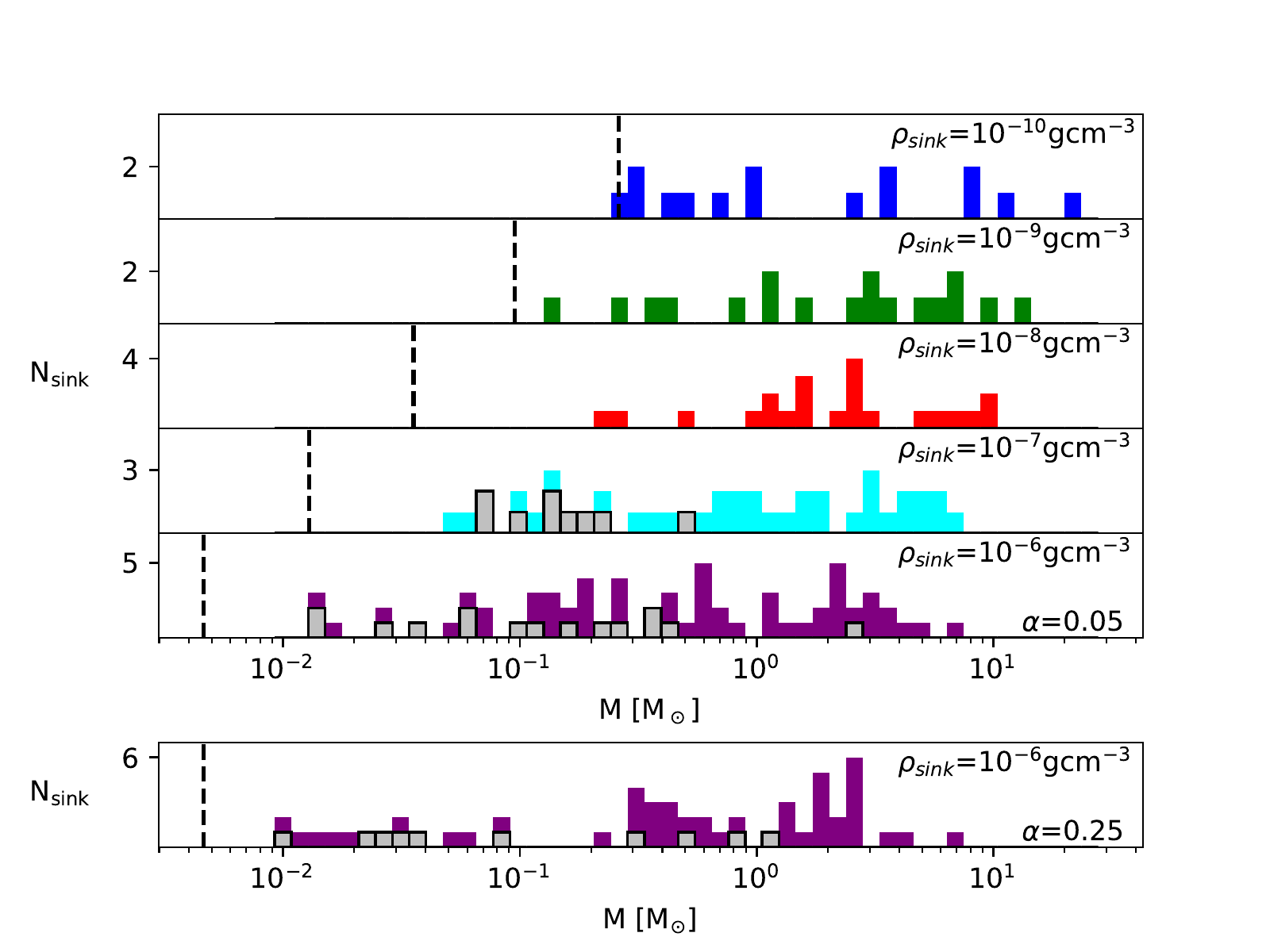}}
    \caption{Initial mass functions at $t\sim$~400 years after the formation of the first sink. The distribution of masses shifts towards lower values with increasing $\rho_{\text{sink}}$. The shaded sections of the IMFs represent sinks ejected from the group, given by the criteria described in Section~\ref{ejec}. The black vertical dashed lines show the Jeans mass at the sink particle creation density. The IMF for the high velocity ($\alpha=0.25$) run is given in a separate panel at the bottom. Note that for $\rho_{\rm sink} = 10^{-10} - 10^{-8} \: {\rm g \: cm^{-3}}$ there are no ejections.}
    \label{fig:IMF}
\end{figure}

\begin{figure}
	 \hbox{\hspace{-1.1cm} \includegraphics[scale=0.63]{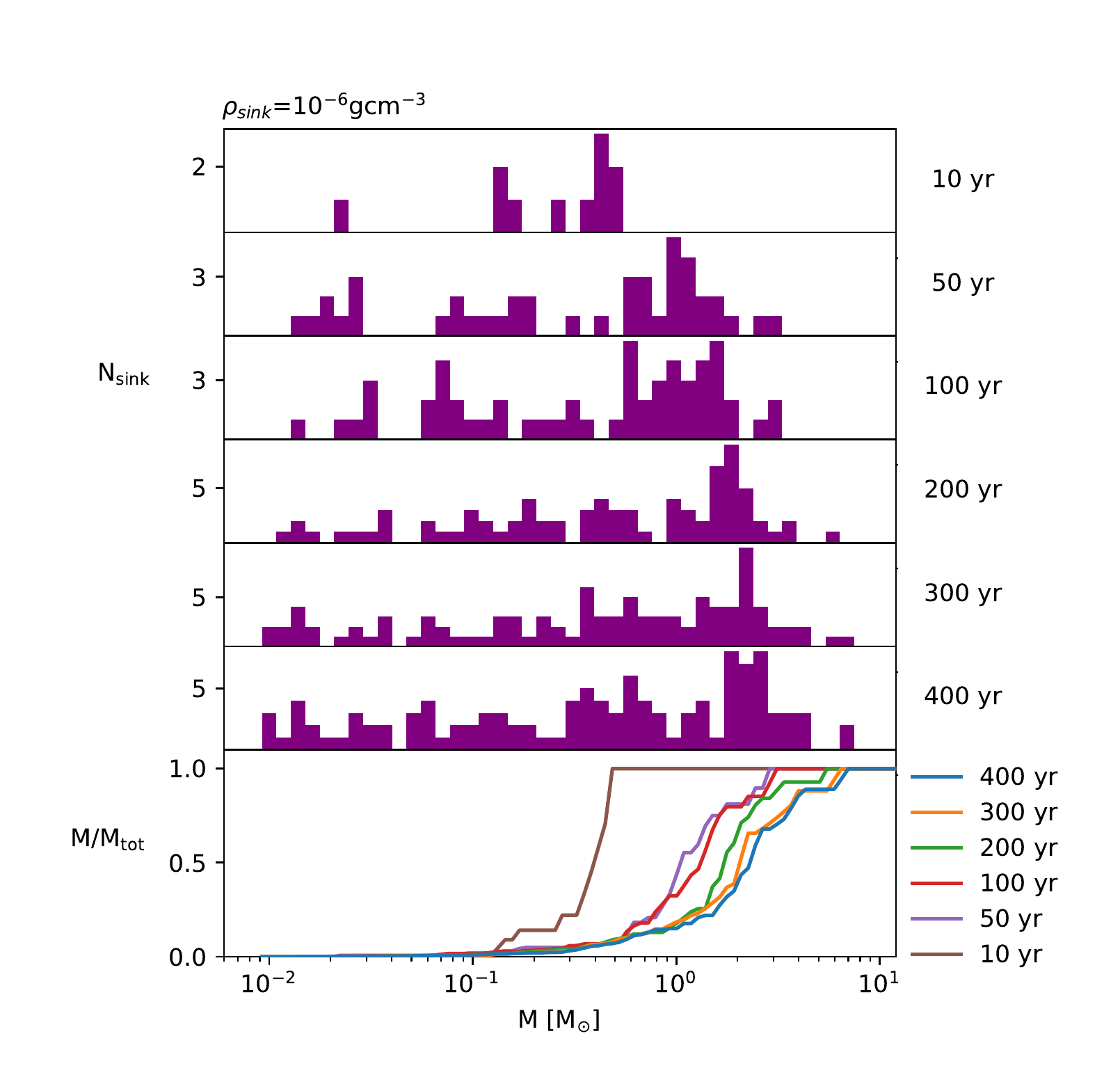}}
    \caption{Evolution of the IMF and cumulative mass function at times 10, 50, 100, 200, 300 and 400 yr after the formation of the fist sink, for $\rho_{\rm sink} = 10^{-6} \: {\rm g \: cm^{-3}}$. The IMFs combine the sinks from the $\alpha=0.25$ and $\alpha=0.05$ runs.}
    \label{fig:IMFtime}
\end{figure}

fig.\ref{fig:IMF} shows the IMF at $\sim$400~yr after the formation of the first sink. The systems are compared at the same time so that their total mass in sinks are the same, as this does not vary with resolution (see Fig.~\ref{fig:sinks}). The IMFs show that the group shifts to a lower mass population with increasing resolution. The shaded regions represent ejections, which are discussed in Section \ref{ejec}. Since accretion onto ejected sinks is halted due to a lack of dense gas, their position on the IMF is fixed. The remaining sinks will continue to accrete, as only a small fraction of the total mass of the cloud has been accreted at 400~yr. 

fig.\ref{fig:IMFtime} shows the time evolution of the IMF and normalised cumulative mass distribution for $\rho_{\rm sink} = 10^{-6} \: {\rm g \: cm^{-3}}$, combining the sinks from the $\alpha=0.05$ and $\alpha=0.25$ runs. By 400 yr, the shape of the IMF appears to be showing convergence.

The present day IMF is bottom-heavy, with most of its mass in low mass stars (e.g. \citealt{Kroupa2019}). The primordial IMFs produced in this study span a similar range to the present day IMF, but most of the mass is in high mass stars. This top-heavy IMF has been produced in many other Pop III studies (e.g.  \citealt{Schneider2006,Susa2014,Stacy2016,Wollenberg2019,Sharda2020}).

\begin{figure}
	 \hbox{\hspace{-0.2cm} \includegraphics[scale=0.7]{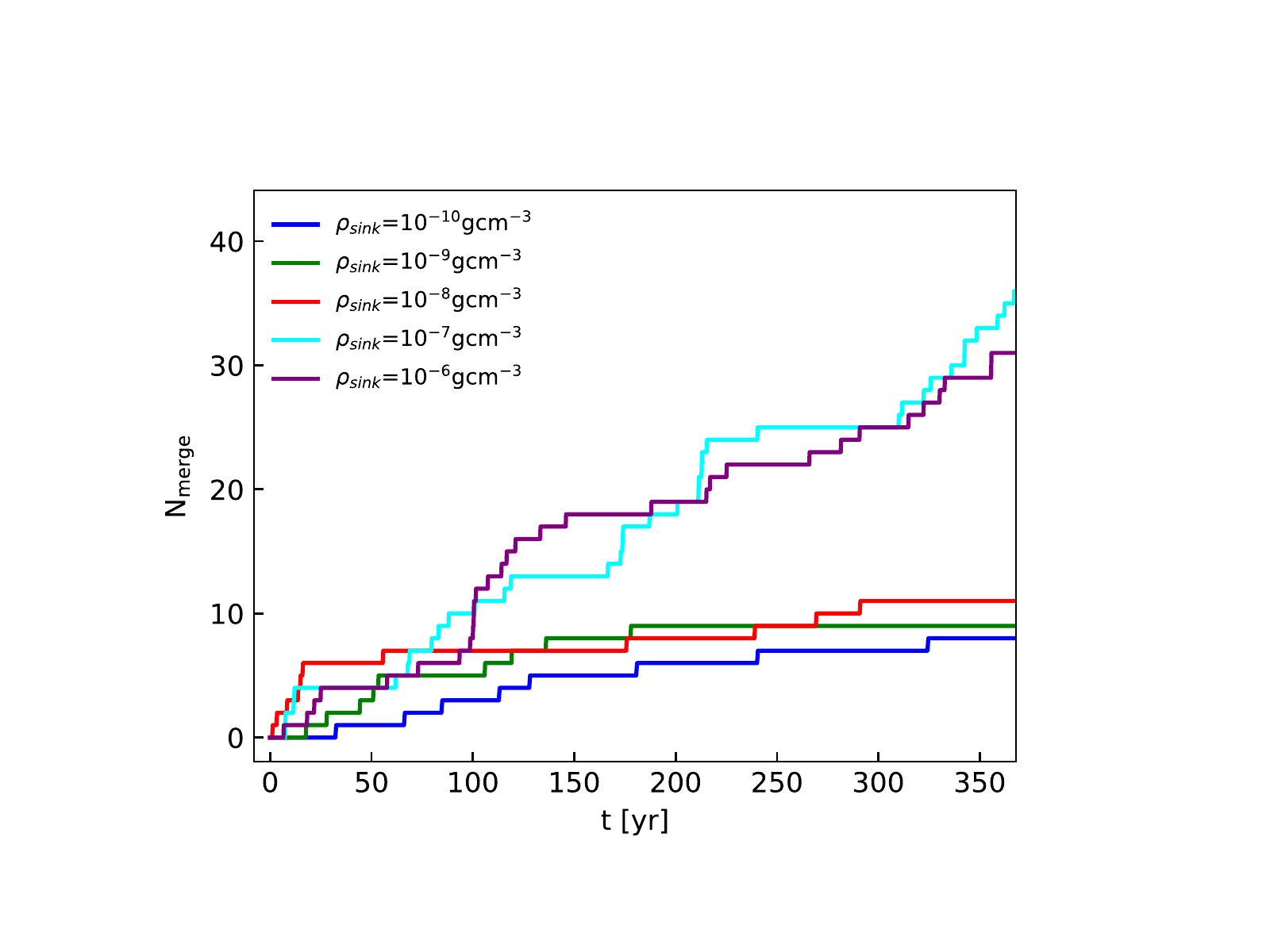}}
    \caption{Cumulative number of sink particle mergers plotted against time across all velocity seeds.}
    \label{fig:mergers}
\end{figure}

\begin{figure*}
	\hbox{\hspace{0.5cm} \includegraphics[scale=0.7]{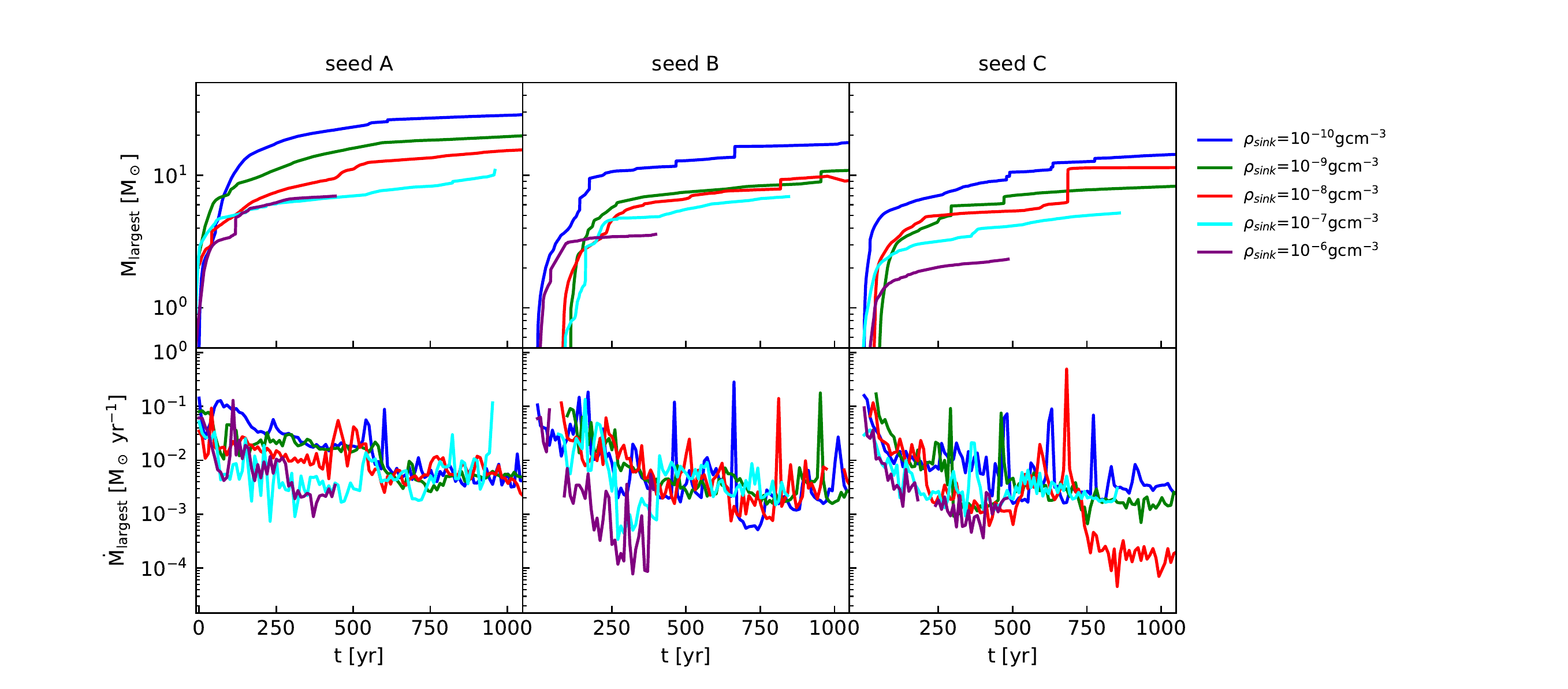}}
    \caption{Time evolution of the mass and accretion rate onto the most massive central sink.}
    \label{fig:masses}
\end{figure*}

More sink particles means more competition for mass accretion, hence lower accretion rates onto the central sink. fig.\ref{fig:masses} shows the mass and accretion rate onto the most massive sink. As the maximum resolution increases, the central sink is starved by the increasingly large number of secondary stars which compete to accrete material that would otherwise fall onto the primary star, resulting in a lower mass central star. This process has been termed `fragmetnation-induced starvation' \citep{Peters2010}. It is not seen perfectly in all of our simulations, as major merger events are capable of boosting the mass of the most massive sink to converge with, or surpass the mass of its lower resolution counterpart. For example, within the seed field A runs, the highest resolution run converges due to a major merger.

\begin{figure}
	 \hbox{\hspace{0cm} \includegraphics[scale=0.6]{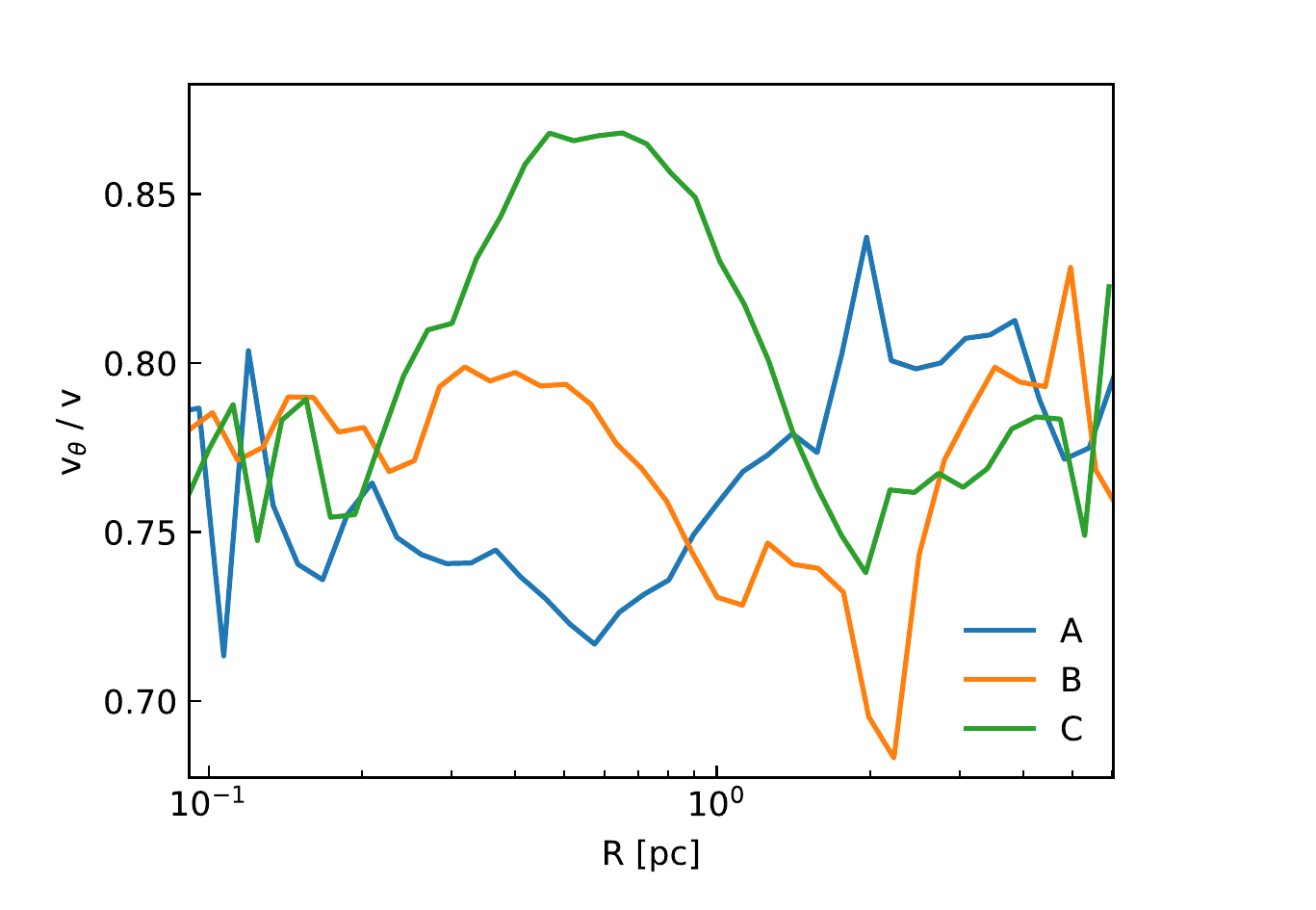}}
    \caption{Radial profiles of the ratio of angular to total velocity in the initial velocity field, around the region where sink particles later form.}
    \label{fig:angular}
\end{figure}

Even within this small set of initial velocity fields, the mass of the largest sink particle shows natural variance. The IMFs of \cite{Wollenberg2019} indicate that turbulent set-ups with lower initial rotation produce higher mass stars. fig.\ref{fig:angular} illustrates the initial ratio of angular to total velocity around the region where the first sink particle later forms. Although we do not explicitly introduce rotation in our simulations, the turbulent velocity field generates some non-zero amount of angular momentum, and we see from the figure that the runs with lower rotational velocity components indeed produce more massive central sink particles. 

\section{Ejections}
\label{sec:ejections}
\label{ejec}
Previous Pop III studies have shown that ejections from the system due to close encounters are common (e.g. \citealt{Smith2011,Greif2012,Machida2013,Wollenberg2019}). Typically once a protostar is ejected from the cloud, it is removed from its accretion source and its ejection mass represents the final mass. We systematically identify ejected sinks by calculating the escape velocity of all sinks, as given by
\begin{equation}
v_{\text{esc}}=\sqrt{\frac{2GM}{R}},
	\label{eq:esc}
\end{equation}
where $R$ is the distance of the sink from the center of mass of the complete set of sinks, which acts as a proxy for the center of the halo\footnote{We choose to use the centre of mass of the sinks rather than that of the gas cells for numerical convenience -- there are far fewer sinks than gas cells. However, as the most massive sinks typically remain close to the centre of the halo, our results should be insensitive to this choice. We also note that the effects of dark matter are negligible at these scales.}. $M$ is the total mass of gas and sinks within a sphere of radius $R$ surrounding this centre of mass. The sink particle is counted as ejected if its velocity exceeds the escape velocity and $R$ exceeds 600~AU. The ejection fraction is given by 
\begin{equation}
f_{\text{eject}}=\frac{\Sigma N_{\text{eject}}(M)} {\Sigma N(M)},
	\label{eq:frac}
\end{equation}
where $N_{\text{eject}}(M)$ and $N(M)$ are the number of ejections and total sinks at mass $M$, and the summations cover a range of stellar masses. Zero metallicity stars with initial masses in the range 0.075--0.8~M$_\odot$ have lifetimes longer than the age of the Universe \citep{Marigo2001}. These stars should therefore be alive today, while stars with masses above 0.8~M$_\odot$ will have depleted their fuel. The ejection histories of the highest resolution runs are shown in the bottom panel of fig.\ref{fig:vels}. The ejection fractions for the highest resolution runs across the 3 seed fields and 2 velocity strengths at $t = 400$~yr after the formation of the first sink are given in Table~\ref{table:2}. We combine all of the sink particles across these 6 highest resolution runs to calculate the mean ejection fractions. We find that sinks of mass below 0.075~M$_\odot$ have a mean ejection fraction of 0.42. These ejected sinks are expected to remain as brown dwarfs for the duration of their lifetimes. The mean ejection fraction for masses between 0.075--0.8~M$_\odot$ was 0.21. These stars are expected to be alive today. Finally, the mean ejection fraction for sinks larger than 0.8~M$_\odot$ was 0.06, which represents stars which have already died by the present day. Table~\ref{table:2} also demonstrates that there is significant run-to-run scatter around these mean values. Ejected sinks are shown by the shaded regions of the IMFs in fig.\ref{fig:IMF}.

The typical halo size is larger than we have simulated in this study (e.g. \citealt{Bromm2002}). However, we argue that since $v \gg v_{\text{esc}}$ for most of the ejected sinks, accretion will only be significant close to the centre of the halo. The Bondi-Hoyle accretion rate is given by 
\begin{equation}
\dot{M} = 4 \pi (GM)^2 \rho / (c_{s}^2 + v^2)^{3/2},
\end{equation}
where $M$ is the sink mass, $\rho$ is the gas density, $c_{\rm s}$ is the sound speed and $v$ is the velocity of the sink relative to the gas. If $v \gg v_{\rm esc}$, then it follows that $v \gg c_{\rm s}$ and that the velocity of the sink will not significantly change as it makes its way out of the halo. Therefore, it follows that
\begin{equation}
\dot{M} \propto \rho.
\end{equation}
Previous work has shown that $\rho \propto r^{-2.2}$ (e.g. \citealt{Yoshida2006}), so for a constant velocity,
\begin{equation}
\dot{M} \propto t^{-2.2}.
\end{equation}
The mass accretion rate therefore drops rapidly with time and the total mass accreted is dominated by accretion close to t = 0. Therefore, the mass that the ejected protostars have accreted in the region that we simulate is the majority of the mass they are ever going to accrete. Hence even if we were to account for the gas on scales larger than a few pc in the minihalo, this would not significantly affect our conclusion regarding
the masses of the ejected protostars.

\begin{table}
	\centering
	\caption{Ejection fractions for brown dwarfs, stars with masses $<$0.8M$_\odot$ and stars with masses $>$0.8M$_\odot$, for the 10$^{-6}$g cm$^{-3}$ runs, at 400~yr after the formation of the first sink.}
	\label{table:2}
	\begin{tabular}{lcccr} 
		\hline
		Seed  & $\alpha$ & $f_{\text{eject}}^{\text{BD}}$ & $f_{\text{eject}}^{<0.8\text{M}_\odot}$ & $f_{\text{eject}}^{>0.8\text{M}_\odot}$\\
		\hline
		
		A & 0.05 & 0.67 & 0.41 & 0 \\
		A & 0.25 & 0.75 & 0 & 0.33 \\
		B & 0.05 & 0.29 & 0.09 & 0.09 \\
		B & 0.25 & 0.22 & 0.40 & 0 \\
		C & 0.05 & 0.67 & 0.22 &  0\\
		C & 0.25 & 0 & 0.25 &  0\\
		All & - & 0.42 & 0.21 & 0.06\\
		
		\hline
	\end{tabular}
\end{table}

\section{Discussion}
\label{sec:discussion}
To the best of our knowledge, the set of simulations presented here is the first systematic study of the impact of varying the sink creation density (and hence the maximum resolution) on the formation of Pop III stars. It is also unusual in that it resolves the collapse up to a very high maximum density ($\rho \sim 10^{-6} \: {\rm g \, cm^{-3}}$ in our highest resolution run) while also following the evolution of the system for hundreds of years after the onset of star formation. Although many previous studies have run for comparable or longer times, they have typically done so with much lower resolution and a far smaller sink creation density \citep[see e.g.\ the overview of previous simulations in][]{Susa2019}.
Conversely, previous studies that have achieved comparable or higher resolution have typically either halted once the first protostar forms \citep[e.g.][]{Yoshida2008} or have been able to run for only a few years after the onset of star formation (e.g.\ \citealt{Greif2012}). The one previous study that we are aware of that combines a similar resolution with a duration of $\sim 100$~yr is the nested grid simulation of \citet{Machida2015}, although this has the disadvantage of only reaching the highest resolution in a small ($\sim 10$~AU) central region in the calculation.

Our resolution study clearly demonstrates that the number of fragments forming within a given time systematically increases as the simulations achieve higher maximum resolution. Following the collapse to higher densities allows us to capture gravitational instability and fragmentation on ever smaller scales, owing to the decrease in $\lambda_{\text{J}}$ with increasing density over the whole range of densities examined here. We have shown that the degree of fragmentation at the highest resolutions is not caused by our choice of $\alpha$, as we recover qualitatively similar results regardless of whether $\alpha = 0.05$ or $\alpha = 0.25$.

Despite increased fragmentation, the total mass of in stars is independent of resolution, consistent with the idea that this is determined primarily by the inflow of gas to the star-forming region, rather than the dynamics within this region. The increase in fragmentation we see at higher resolution therefore shifts the system from containing a few high mass stars at low resolutions to having many stars with lower masses in high resolution runs. The number of sinks formed did not converge, despite the fact that at the highest densities, the gas is optically thick and unable to cool radiatively. H$_{2}$ collisional dissociation enables the gas to dissipate thermal energy and remain approximately isothermal, allowing it to continue to fragment. This suggests that the degree of fragmentation will continue to increase with resolution until H$_{2}$ is fully dissociated at densities $\rho \sim$10$^{-4}$~g cm$^{-3}$. Since most Pop III studies have used sink particle creation densities lower than the maximum tested here (e.g. \citealt{Greif2011,Smith2011,Clark2011,Susa2014,Stacy2016,Wollenberg2019,Sharda2020}), we expect that the IMFs produced by these studies have underestimated the number of stars formed and overestimated the mass of the stars.

The uncertainty in the IMF also affects the amount of ionising radiation produced by the system. An IMF dominated by high mass stars will output significantly more ionising photons than an IMF dominated by low mass stars. Once Pop III stars start emitting UV radiation, it can drastically affect the star-forming environment. Photons with energies above the Lyman limit ionise the neutral hydrogen around the star to form an H$\,${\sc ii} region  known as a Str\"{o}mgren sphere (\citealt{Stromgren1939,McCullough2000}). Lower energy UV photons in the Lyman-Werner (LW) band escape the H$\,${\sc ii} region and are able to excite and subsequently dissociate molecular H$_{2}$, the main coolant in primordial star formation. Although it is currently unclear when and how effectively both forms of radiation escape from the immediate vicinity of the massive Pop III stars \citep{Jaura2021}, a number of simulations have outlined the effects that it can have on the gas within Pop III star-forming minihalos. For example, simulations by \cite{Stacy2012} show that ionisation of neutral hydrogen and photo-dissociation of H$_2$ by LW radiation causes the accretion rate onto surrounding protostars to reduce by an order of magnitude by the time the central star reaches 20~M$_{\odot}$. The ionising output of the first stars can also effect the next generation of stars. For instance, \cite{Hirano2015} showed that stars forming in the radiation field of the first generation of stars have stunted accretion rates and masses, which in turn lowers their output of ionising radiation (see also e.g. \citealt{Haiman1997,Omukai2003,Reed2005}).


Complicating the picture here is the influence of the change in the accretion rates that occurs as we change the resolution. At higher resolution, we form more fragments and hence fragmentation-induced starvation of the most massive protostars becomes increasingly pronounced, as illustrated in fig.\ref{fig:masses}. Protostars with accretion rates $\sim 10^{-2} \: {\rm M_{\odot} \: yr^{-1}}$ or higher will be highly inflated, with large radii and low photospheric temperatures \citep[see e.g.][]{Omukai2003, Hirano2014}, and will only approach the standard zero-age main sequence once their accretion rates drop below this value. Protostars formed in simulations with low $\rho_{\rm sink}$ will therefore reach the main sequence more slowly than they should, delaying the time at which they become efficient emitters of ionising photons.\footnote{In Appendix~\ref{B}, we estimate the total ionising luminosity generated by accretion onto the protostars and show that this is negligible compared to the ionising luminosity of even a single massive main sequence star.} In practice, fig.\ref{fig:masses} suggests that this effect is less important than the substantial change in the IMF that we see with increasing resolution, and that the overall impact of increased resolution is to delay the time at which photoionisation feedback becomes important.

fig.\ref{fig:mergers} shows the cumulative number of mergers as a function of time for all $\rho_{\text{sink}}$. Generally the number of mergers increases with the maximum resolution, which is to be expected as more sink particles form. However, despite the increase in $N_{\text{sink}}$ from 10$^{-7}$ to 10$^{-6}$g cm$^{-3}$, the number of mergers appears to converge. This coincides with the resolution where sink particle ejections are first seen in our simulations. We speculate that the merging behaviour converges as a consequence of the loss of sinks from the system via ejections.

Finally, our simulations demonstrate that many low mass stars will be ejected from a Pop III star-forming minihalo as it evolves, with a substantial fraction having masses that allow them to still be alive today. While accretion of metals from interstellar objects (ISOs) may disguise low mass ejectors as Pop II stars \citep{Tanikawa2018}, most studies indicate that a stellar wind and radiation can blow away interstellar particles before reaching the stellar surface (\citealt{Tanaka2017,Johnson2011}). This would leave these low mass Pop III ejectors as pristine, metal free stars. As low-mass stars below $\sim 0.8\,$M$_{\odot}$ will have survived until the present days \cite[][]{kippenhahn2012a}, these objects should be directly detectable in current and future stellar archeological surveys \cite[see e.g.][]{beers2005a, frebel2015, starkenburg2017}. Even non-detections allow us to constrain the low-mass end of the Pop III IMF. For example, \citet{hartwig2015c} and \citet{magg2019} computed the expected numbers of low-mass Pop III stars in the Galactic halo based on semi-analytic models of the assembly history of the Milky Way, and estimated the minimum sample size needed to exclude the existence of genuine low-mass Pop III stars in our Galaxy. For a similar discussion, see also \citet{salvadori2007a, salvadori2010a} and  \citet{tumlinson2006a, tumlinson2010a}. This approach can be extended to include the satellite galaxies to the Milky Way \citep[e.g.][]{magg2018}. 

Although protostellar ejections have been found to occur in many previous simulations of Pop III star formation, \cite{Greif2012} raised the concern that this may be a numerical artifact, a consequence of not allowing sink particles to merge. We have introduced sink mergers into {\sc arepo} to remedy this situation and have proved that ejections indeed continue to occur when merging is enabled, suggesting that this is a real physical effect and not an artifact.

\section{Caveats} 
\label{sec:caveats}
Our initial conditions represent idealised Bonnor-Ebert sphere density profiles in randomised turbulent velocity fields, in the absence of a live dark matter gravitational potential. Although we expect the influence of the dark matter halo to be small on the scales we simulate, it is nevertheless the case that a more realistic approach would be to isolate a single dark matter halo and perform star formation simulations within it. However, this is not necessary to assess the effects of resolution on the fragmentation behaviour of primordial gas, as is the focus here.

A more important limitation of our current study (albeit one shared by most previous work on Pop III star formation) is the fact that we have examined only non-magnetised collapses. Weak magnetic seed fields are believed to have existed in the early Universe, which were amplified by the small-scale turbulent dynamo, resulting in a magnetic field with a $k^{3/2}$ power spectrum \citep{Sur2010, Schleicher2010, Schober2015}. Present day star formation simulations show that uniform magnetic fields support discs against fragmentation (e.g.\ \citealt{Hennebelle2008, Seifried2011, Burzle2011}) and evidence is starting to accumulate that the small-scale primordial fields can also reduce fragmentation \citep{Sharda2020}. Magnetic flux-freezing allows the field to become stronger as the maximum density of the gas is increased, which may counteract the increase in gravitational instability as the maximum resolution is increased. The ability of magnetic fields to achieve numerical convergence in $N_{\text{sink}}$ and prevent increasing fragmentation-induced starvation will be assessed in a future paper.

Another caveat is the fact that in our treatment of sink particle mergers, we have assumed that the merging cross-section is fixed, and set by the accretion radius of the sink particle, which in turn is set by the Jeans length at the sink formation density. In reality, the merging scale will depend on the physical size of the protostar and likely increases as the protostar gains mass.  
Finally, these simulations only included accretion luminosity feedback and not the radiative feedback produced by the stars once they reach the main sequence. Although stellar luminosity has been shown to prevent accretion onto the host star and to suppress star formation in the surrounding cloud \citep{Stacy2016}, its absence from our simulations is unimportant, since the protostars formed in our simulations are still in the early, adiabatic phase of their evolution and will not emit significant radiation. 

\section{Conclusions}
\label{sec:conclusion}
In a Population III setting, the effect of varying the sink particle creation density was investigated. A turbulent cloud collapse resulting in periodic bursts of star formation was repeated for sink creation densities $\rho_{\text{sink}}$ spanning the range $10^{-10}$-$10^{-6}$g cm$^{-3}$. As $\rho_{\text{sink}}$ increased, larger numbers of sinks were formed without converging within the range tested. The total mass in sinks was independent of the sink parameters used, resulting in an IMF that shifts towards lower mass stars with higher sink creation density. We also find that a significant number of low mass protostars get ejected from the halo center by stellar dynamical processes. Combining the 6 iterations of the highest resolution case resulted in an ejection fraction of 0.21 for low mass stars capable of surviving to this day (0.075 - 0.8 M$_\odot$), and 0.42 for brown dwarf stars. Our results suggest that the number of sinks will continue to grow with increasing $\rho_{\text{sink}}$, until H$_{2}$ is fully dissociated and the collapse becomes almost adiabatic at 10$^{-4}$g cm$^{-3}$. More sinks owing to higher $\rho_{\text{sink}}$ also caused increased fragmentation-induced starvation of the most massive sink in the system, which will lower the final stellar mass and subsequent ionising radiation output of the star. Since Population III studies are yet to reach densities of 10$^{-4}$g cm$^{-3}$, these results show that the primordial IMF has thus far overestimated the mass of stars and underestimated the number of stars forming per halo. We predict a significant number of free-floating genuine low-mass Pop III stars in the Milky Way and its satellites, which could be the target of current and future archaeological surveys.

\section*{Acknowledgements}
This work used the DiRAC@Durham facility managed by the Institute for Computational Cosmology on behalf of the STFC DiRAC HPC Facility (www.dirac.ac.uk). The equipment was funded by BEIS capital funding via STFC capital grants ST/P002293/1, ST/R002371/1 and ST/S002502/1, Durham University and STFC operations grant ST/R000832/1. DiRAC is part of the National e-Infrastructure.

The authors gratefully acknowledge the Gauss Centre for Supercomputing e.V. (www.gauss-centre.eu) for supporting this project by providing computing time on the GCS Supercomputer SuperMUC at Leibniz Supercomputing Centre (www.lrz.de).

We also acknowledge the support of the Supercomputing Wales project, which is part-funded by the European Regional Development Fund (ERDF) via Welsh Government.

RSK and SCOG acknowledge support from the Deutsche Forschungsgemeinschaft (DFG, German Research Foundation) via the collaborative research center (SFB 881, Project-ID 138713538) ``The Milky Way System'' (subprojects A1, B1, B2 and B8). They also acknowledge support from the Heidelberg Cluster of Excellence ``STRUCTURES'' in the framework of Germany’s Excellence Strategy (grant EXC-2181/1, Project-ID 390900948) and from the European Research Council (ERC) via the ERC Synergy Grant ``ECOGAL'' (grant 855130).

We thank our referee Dr Christopher F. McKee, for recommendations that improved the manuscript.

\section{Data Availability}
The data underlying this article will be shared on reasonable request to the corresponding author.



\bibliographystyle{mnras}

\bibliography{references.bib} 



\appendix
\section{Seed fields B \& C}
\label{A}
\begin{figure*}
	\hbox{\hspace{-0.6cm} \includegraphics[scale=0.7]{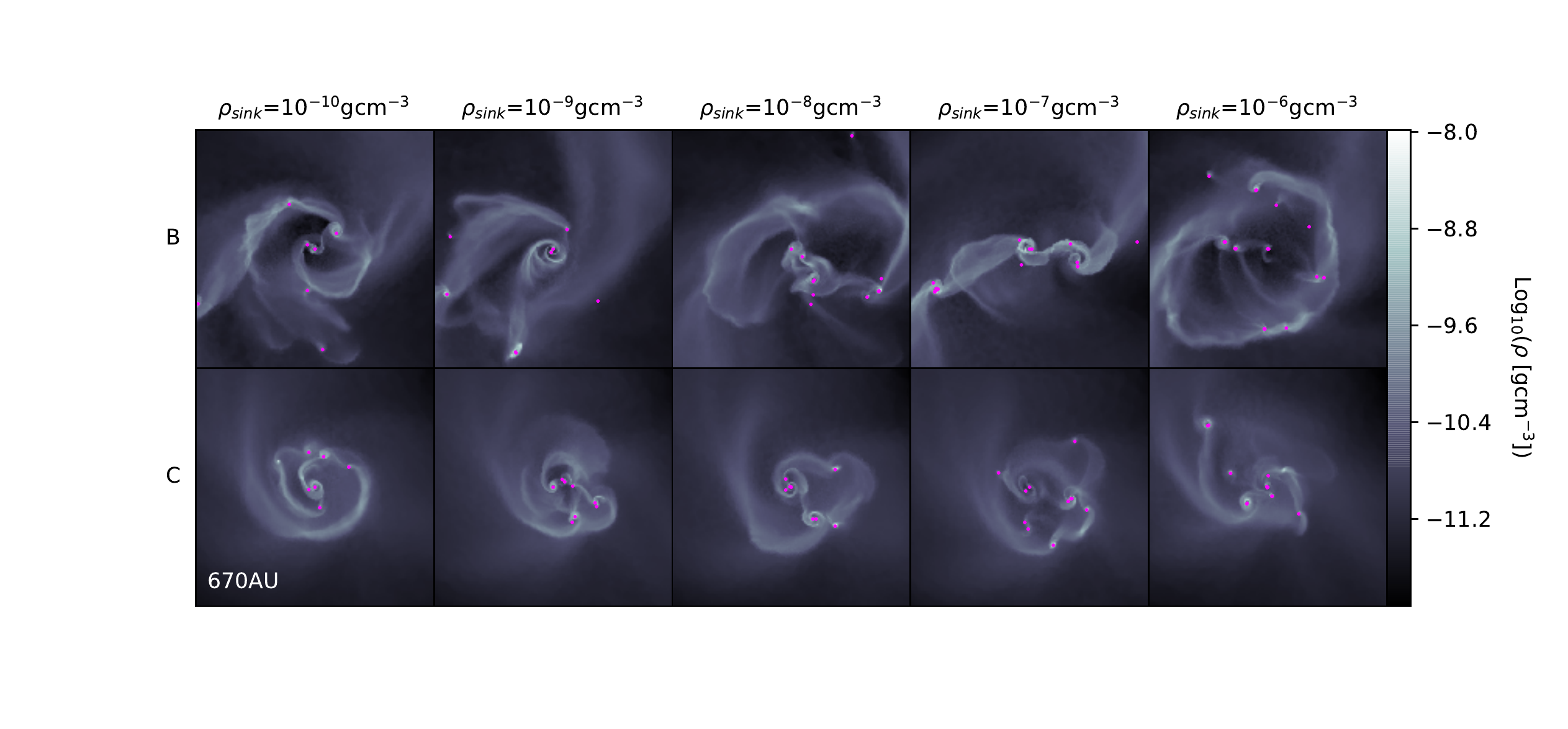}}
    \caption{Initial velocity fields B and C. Inner 670~AU of the {\sc Arepo} unstructured density distributions projected onto uniform 500$^3$ grid cubes, flattened by summing the density over the $z$-axis. Snapshots taken at 400 years after the formation of the first sink particle, for $\rho_{\text{sink}}$=$10^{-10}$-$10^{-6}$g cm$^{-3}$. Sink particles are shown as magenta dots. }
    \label{fig:grid2}
\end{figure*}

We have opted to only show the time evolution of the structure of the system for seed field A. For seed fields B and C, we show only a comparison of the systems at 400~yr after the formation of the first sink for each resolution case, in fig.\ref{fig:grid2}. The inner 670~AU of the {\sc Arepo} unstructured density distributions were projected onto uniform 500$^3$ grid cubes and were flattened by summing the density over the $z$-axis. It is clear from the figure that although there are differences in the details of the fragmentation in each case, the overall evolution is qualitatively similar for all three seed fields. 

\section{Accretion luminosity radiation}
\label{B}

\begin{figure*}
	\hbox{\hspace{0.5cm} \includegraphics[scale=0.7]{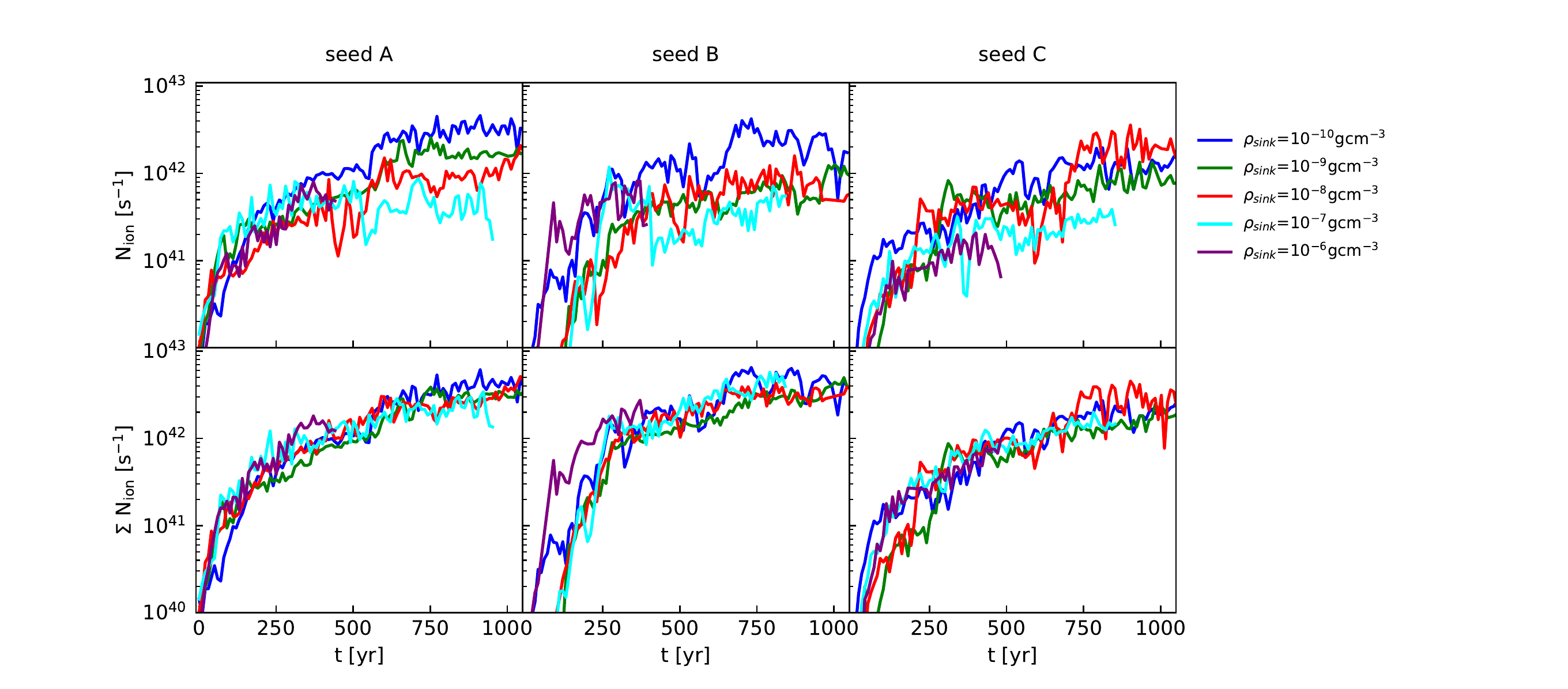}}
    \caption{Ionising photons per second generated by accretion onto the most massive sink in the system (top) and for all sinks combined (bottom). }
    \label{fig:Nion}
\end{figure*}

fig.\ref{fig:masses} shows that the accretion rate onto the largest sink is initially $\sim$10$^{-1}$M$_{\odot}$yr$^{-1}$ for all resolutions and all three seed fields, but that it drops rapidly to $\sim$10$^{-2}$M$_{\odot}$yr$^{-1}$ or below within a few hundred years. Given the high accretion rates, it is also interesting to ask whether the contribution made to the ionising flux by accretion onto the protostars is ever significant. To investigate this, we estimate the number of ionising photons emitted per second due to accretion using the following prescription. We first calculate the accretion luminosity of each protostar:
\begin{equation}
L_{\text{acc}}=\frac{GM\dot M}{R_{\star}},
\label{eq:lum}
\end{equation}
where $M$ and $\dot M$ are the mass of, and accretion rate onto the protostar, respectively. We take $R_{\star}$ from the mass-radius relationship given in \cite{Hosokawa2009} as
\begin{equation}
R=26R_\odot \left(\frac{M}{M_\odot}\right)^{0.27} \left(\frac{\dot M}{10^{-3}M_\odot \text{yr}^{-1}}\right)^{0.41}.
\label{eq:radius}
\end{equation}
The effective temperature $T_{\rm eff}$ of the protostar can be estimated by equating the accretion luminosity to the total power radiated from a black body via the Stefan–Boltzmann's law:
\begin{equation}
\frac{L_{\text{acc}}}{ 4\pi R_\star^2}= \sigma T_{\rm eff}^4,
\label{eq:temp}
\end{equation}
where $\sigma$ is the Stefan–Boltzmann constant. Using $T_{\rm eff}$, the total ionising energy per second can be estimated by integrating the Plank function for frequencies higher than the Lyman limit (3.28$\times$10$^{15}$~Hz). The number of ionising photons emitted per second then follows as:
\begin{equation}
\dot N_{\text{ion}} =4\pi  \int_{\text{UV}} \frac{B_v(T) 4\pi R_\star^2 }{hv} dv,
\end{equation}
where $B_v(T)$ is the Plank function, $h$ is Plank's constant and $v$ is the frequency of the the photons. The number of ionising photons per second are shown in fig.\ref{fig:Nion}. The top panel shows that the lower mass stars formed in higher resolution runs produce less ionising accretion radiation as a result of the fragmentation-induced starvation. The bottom panel shows that the total accretion ionising radiation from all stars in the system is unaffected by the resolution used. This can be explained by substituting equations \ref{eq:lum} and \ref{eq:radius} into \ref{eq:temp} to obtain 

\begin{equation}
T \propto \frac{M^{0.0475}}{\dot M ^{0.575}},
\end{equation}
which shows that for a given mass, higher temperatures are obtained from lower accretion rates due to a smaller radius. This allows the low mass stars in the higher resolution runs to boost their ionising radiation output compared to sinks of the same mass in mass in lower resolution runs. This combined with the greater number of sinks produced in higher resolution runs explains how the total ionising radiation is independent of the resolution used and subsequent fragmentation-induced starvation.


\bsp	
\label{lastpage}

\end{document}